\documentclass[runningheads]{llncs}
\usepackage{graphicx}
\usepackage{amsmath}
\usepackage{subcaption}
\usepackage{multicol}
\usepackage{multirow}

\begin{document}
\title{Efficient High-Resolution Image-to-Image Translation using Multi-Scale Gradient U-Net}

\author{Kumarapu Laxman \and
Shiv Ram Dubey \and
Baddam Kalyan \and 
Satya Raj Vineel Kojjarapu}

\authorrunning{Laxman Kumarapu et al.}
\titlerunning{MSG U-Net}

\institute{Computer Vision Group\\
Indian Institute of Information Technology, Sri City, Chittoor, Andhra Pradesh, India
\email{\{laxman.k17, srdubey, kalyan.b17, satyarajvineel.k17\}@iiits.in}}

\maketitle

\begin{abstract}
Recently, Conditional Generative Adversarial Network (Conditional GAN) have shown very promising performance in several image-to-image translation applications. However, the uses of these conditional GANs are quite limited to low-resolution images, such as $256 \times 256$. The Pix2Pix-HD is a recent attempt to utilize the conditional GAN for high-resolution image synthesis. 
In this paper, we propose a Multi-Scale Gradient based U-Net (MSG U-Net) model for high-resolution image-to-image translation up to $2048 \times 1024$ resolution. 
The proposed model is trained by allowing the flow of gradients from multiple-discriminators to a single generator at multiple scales. 
The proposed MSG U-Net architecture leads to photo-realistic high-resolution image-to-image translation. 
Moreover, the proposed model is computationally efficient as compared to the Pix2Pix-HD with an improvement in the inference time nearly by $2.5$ times. We provide the code of MSG U-Net model at \url{https://github.com/laxmaniron/MSG-U-Net}.

\end{abstract}
\section{Introduction}
\label{sec:intro}
Recently, image-to-image translation has gained huge popularity after the success of Generative Adversarial Networks (GANs) \cite{goodfellow2014generative} in a wide range of image processing and computer vision applications \cite{babu2020csgan}, \cite{patil2020end}, \cite{babu2020pcsgan}, \cite{nema2020rescuenet}. In image-to-image translation, an image in a domain is generally transformed into the corresponding image in some other domain, such as translating semantically segmented label images into RGB images, aerial photos to maps, sketches to real faces, and indoor segmented images to real images. 

In 2014, GAN was proposed by Goodfellow et al. \cite{goodfellow2014generative} to synthesize photo-realistic images with the help of generator and discriminator networks. In 2017, the GAN was extended for image-to-image translation in form of conditional GAN (i.e., Pix2Pix model) \cite{C2} which became very popular. Motivated from the conditional GAN, several variants of GANs have been developed for image-to-image translation \cite{CycleGAN}, \cite{park2020contrastive}, \cite{SPAGAN}. However, these GAN models are generally not designed for the high-resolution image-to-image translations. Moreover, these models can only produce images of good quality up to $256 \times 256$ resolution.

To improve the quality of generated images, Perceptual Adversarial Networks (PAN) for Image-to-Image Transformation \cite{C3} were proposed which uses perceptual losses from pre-trained image classification networks such as VGG-19 \cite{C7}. However, its output resolution is also limited to $256 \times 256$. In an another attempt, a progressive GAN \cite{C8} is proposed to train the generator and discriminator networks with first low-resolution images then high-resolution images, which poses an additional burden on the training. In 2018, Ting-Chun Wang et al. \cite{C1} proposed the Pix2Pix-HD model for high-resolution image synthesis and semantic manipulation with the help of conditional GANs. The Pix2Pix-HD \cite{C1} performs well for high-resolution image-to-image translation (up to $2048 \times 1024$) as well. However, it is computational resource hungry (the complexity of Pix2Pix-HD is about 3.32 TeraFlops) and requires huge memory (at least 8GB of GPU VRAM) even during the inference time. 
Recently, multi-scale gradient for GAN (MSG-GAN) \cite{C5} is introduced for the high-resolution image generation by utilizing the gradient information at different scales. However, the power of multi-scale gradient is not yet utilized for the image-to-image translation.
In this paper, we tackle these issues of high-resolution image-to-image translation with the help of the proposed efficient Multi-scale Gradient U-Net inspired from MSG-GAN \cite{C5} which requires significantly less computational resources and generates the high quality images.

Following are the main contributions of this paper:
\begin{itemize}
    \item We propose a novel Multi-Scale Gradient U-Net (MSG U-Net) based GAN model for high-resolution image-to-image translation (up to $2048 \times 1024$) which is inspired from the MSG-GAN \cite{C5} architecture. 
    \item The proposed MSG U-Net is very efficient as compared to the state-of-the-art GANs for high-resolution image-to-image translation. MSG U-Net requires half the memory (4GB of GPU VRAM) and computational resources (the complexity of MSG U-Net is about 1.28 TeraFlops) during inference time as compared to Pix2Pix-HD \cite{C1} while retaining the same level of details in the output images. 
    % both qualitatively and quantitatively.
    % \item  The quantitative results compared to other GAN architectures is shown in section \ref{sec:experimental_results}.
\end{itemize}

\begin{figure*}[!ht]
\centering
  \includegraphics[width=\textwidth]{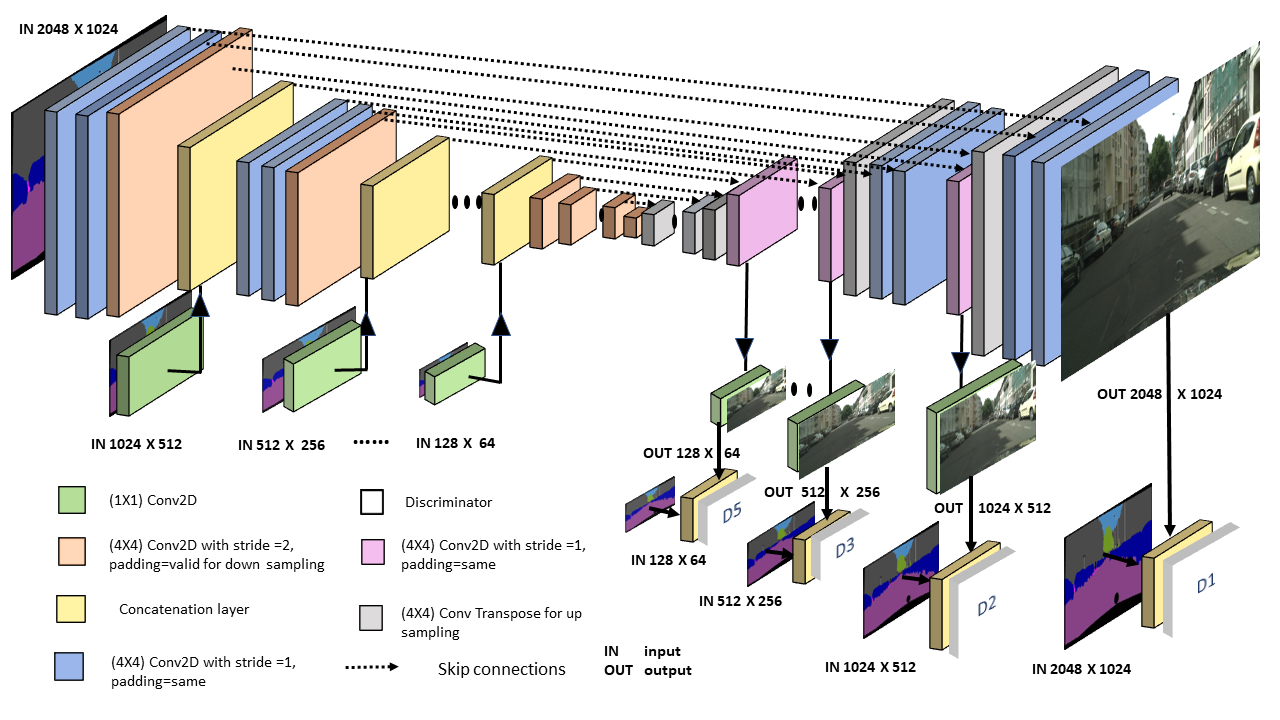}
  \vspace{-0.5cm}
  \caption{The above figure shows the Network architecture of our proposed Multi-Scale Gradient U-Net (MSG U-Net) GAN model. The input from the source domain is provided to the network at various resolutions ranging from $128 \times 64$ to $2048 \times 1024$. Similarly, the generator network generates the output images in the target domain at various resolutions ranging from $128 \times 64$ to $2048 \times 1024$. The output image generated at a particular resolution is concatenated with the input image from the source domain at the same resolution and fed into the corresponding discriminator similar to Pix2Pix Conditional GAN \cite{C2}. The network uses multiple discriminators, i.e., one discriminator for each resolution.}
  \label{fig:fig1}
\end{figure*}

\section{Proposed MSG U-Net GAN Model}
\label{sec:proposedmethod}
In this section, we present the generator and discriminator architectures along with the objective function of the proposed MSG U-Net model.

\subsection{Generator Architecture}
\label{ssec:genarchitecture}
MSG-GANs \cite{C5} generate high-resolution images up to $1024 \times 1024$ from latent vectors. It avoids the vanishing gradient problem faced by very deep networks by allowing the flow of gradients from the discriminator to the generator at multiple scales. Inspired by this work, we propose a new generator architecture multi-scale gradients for U-Net (MSG U-Net) for high-resolution image-to-image translation (upto $2048 \times 1024$ resolution) as shown in Fig. \ref{fig:fig1}.

\subsubsection{Encoder Part of MSG U-Net}
\label{sssec:encoder}
Let the input image from the source domain be of resolution $2048 \times 1024$ fed to the encoder part of the network. The input image is passed through two convolutional blocks of kernel size $4\times4$ and stride $1$. Each convolutional block consists of  Convolution-BatchNorm-LeakyReLu layers. Then the resolution is reduced to $1024 \times 512$ by using a convolutional block of kernel size $4\times4$ with stride as $2$. Then the output of the convolutional block is concatenated with the resized source image ($1024 \times 512$), passed through a convolutional block having kernel size $1 \times 1$. In this way, we form the encoder part of the generator by providing input from the source domain at various scales from $2048 \times 1024$ to $128 \times 64$. After $128 \times 64$ input block, the encoder follows normal U-net architecture, and the resolution of convolutional block is reduced to $8 \times 4$. Providing input at different scales at different stages to the encoder leads to the learning of important features such as overall shape information from the low-resolution and detailed texture information from the high-resolution image.
% way of feeding same image at various scales has advantages as we are feeding low-resolution source image which consists of overall shape information, and high-resolution source image which consists of detailed texture to encoder, this will give encoder more information about the image.

\subsubsection{Decoder Part of MSG U-Net}
\label{sssec:decoder}
The output of the encoder from the $8 \times 4$ convolutional block is considered as the input to the decoder. First, the decoder up-samples to $8 \times 16$ using the transposed convolution block (TransposedConvolution-BatchNorm-LeakyRelu). The $8 \times 16$ output from the transposed convolutional block is added to the output from the corresponding $8 \times 16$ encoder block using skip connections followed by a convolutional block with stride $1$. In this manner, the decoder upsamples upto $2048 \times 1024$ using transposed convolutional blocks. In addition to this, the decoder has branching at different layers to generate output images in the target domain from $128 \times 64$ to $2048 \times 1024$ resolutions simultaneously. 
% The target images generated are of different scales/resolution.

The process of generating outputs at intermediate layers in the generator network is inspired from MSG-GANs \cite{C5} and GoogLeNet \cite{C6}. It helps the network to alleviate the problem of vanishing gradients. 
% Thus, the utilization of generating outputs from intermediate layers in the proposed method solves the problem of vanishing gradients for the deeper U-Net. 
As the discriminators not only take generator's final output as input but also the intermediate outputs, the gradients can flow from discriminators to the intermediate layers of the generator directly. This increases the stability during training and solves the problem of vanishing gradients when the U-Net architecture is very deep.

\begin{figure}[!t]
\centering
\includegraphics[width=0.49\columnwidth]{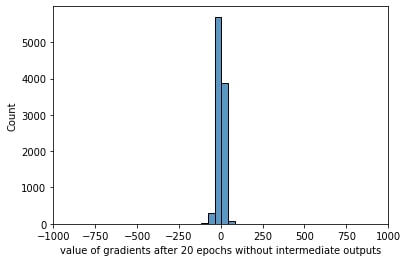}
\includegraphics[width=0.49\columnwidth]{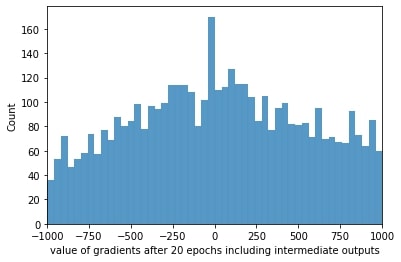}
\caption{The gradient distribution after $20$ epochs of training when (left) the output at intermediate layers is not used, and (right) the output at intermediate layers is used.}
\label{fig:bad_grad}
\end{figure}

To illustrate the impact of having output of different resolutions at intermediate layers, we train the MSG U-Net with and without the intermediate output images for $20$ epochs and show the gradient distribution in Fig. \ref{fig:bad_grad}. It can be noticed that the gradients are very close to zero due to the gradient diminishing problem when output at intermediate layers  are not used. Whereas the gradient is distributed better when the output at intermediate layers are used. 

\subsection{Discriminator Architecture}
\label{ssec:discarchitecture}
We use a modified version of multi-scale discriminators proposed in \cite{C1} while training. As our generator produces target images at different scales, we scale the source image to various scales and concatenate with generated output at corresponding scale and feed it to the discriminator as a fake sample(0). We also concatenate the source and actual target image at corresponding scale and feed it to discriminator as a real sample(1). Images at different resolutions are fed into different discriminators. However, the architecture of discriminator is same at all scales. We use the same patch discriminator as used in Pix2Pix \cite{C2}. We found that multiple discriminators lead to better stability during training than using single discriminator for all scales described in MSG-GAN \cite{C5}.

\begin{figure}[!t]
\centering
\includegraphics[width=0.8\columnwidth]{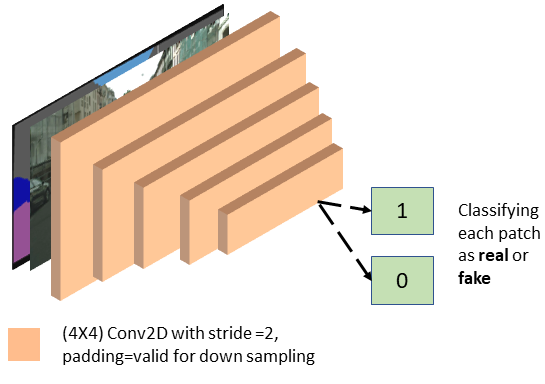}
\caption{Network architecture of discriminator, each discriminator handles image at one resolution/scale.}
\label{fig:disc_architecture}
\end{figure}

\subsection{Loss Function}
\label{ssec:lossfunction}
We use the improved adversarial loss proposed in Pix2Pix-HD \cite{C1} along with perceptual loss using VGG-19 \cite{C7} as our loss function for generator and all discriminators. Let $G$ be the  generator and $D_{1}, D_{2}, ..., D_{n}$ be $n$ discriminators. Let $D_{k}^{(i)}(X,Y)$ be the output of the $i^{th}$ layer of discriminator $D_{k}$ when fed with input images \{$X,Y$\}. Let $E_{i}$ be the total no. of elements in the $i^{th}$ layer of discriminator and $P$ be total number of layers in discriminator. Consider $x_{k}$ as the source image and $y_{k}$ as the corresponding target image at  resolution $k$ with $z_{k}$ as the corresponding generated output. For perceptual loss, assume $F_{k}$ as the pre-trained VGG-19 \cite{C7} with frozen weights handling inputs at resolution $k$, $U$ as the number of convolution layers in VGG-19 and $F_{k}^{(i)}(X)$ be the output of the $i^{th}$ layer of VGG-19 network $F_{k} $ when fed with input image $X$. Consider $V_{i}$ be the total no. of elements in $i^{th}$ layer of VGG-19. The final objective/loss function $L_{total}$ of the proposed MSG U-Net model can be written as:
\begin{equation}
\begin{aligned}
\label{eq:lossfunc}
  L_{total} = \max_{G}\min_{D_{1},D_{2} ... D_{n}}\sum_{k=1}^{n} L_{GAN}(G,D_{k}) + \\ \alpha \sum_{k=1}^{n}\sum_{i=1}^{P}\frac{1}{E_{i}}[||D_{k}^{(i)}(x_{k},y_{k})-D_{k}^{(i)}(x_{k},z_{k})||_{1}]+\\
  \beta \sum_{k=1}^{n} \sum_{i=1}^{U}\frac{1}{V_{i}}[||F_{k}^{(i)}(y_{k})-F_{k}^{(i)}(z_{k})||_{1}]
\end{aligned}
\end{equation}
% Here in equation \ref{eq:lossfunc} $  
where $L_{GAN}$ denotes the adversarial loss and $\alpha$ and $\beta$ are multiplication factor for feature loss \cite{C1} and perceptual loss \cite{C3}, respectively.

\begin{table*}[!t]
\centering
\small
\caption{The results comparison of the proposed MSG U-net with state-of-the-art image-to-image translation GANs on Cityscapes \cite{Cordts2016Cityscapes}, NYU Indoor RGBD \cite{Silberman:ECCV12} and Aerial Images to Maps \cite{C2} datasets.}
\begin{tabular}{|c |c c c| c c c| c c c|}  
 \hline
 Model & \multicolumn{3}{|c|}{Cityscapes } & \multicolumn{3}{|c|}{NYU Indoor RGBD } & \multicolumn{3}{|c|}{Aerial Images to Maps }\\
 \cline{2-10}
  & PSNR  & SSIM & VIF & PSNR  & SSIM & VIF & PSNR  & SSIM & VIF\\ [0.5ex]
 \hline
 Pix2Pix $(256\times256)$ \cite{C2} &  15.74 & 0.42 & 0.05 & 18.08 & 0.45 & 0.07 & 26.20 & 0.64 & 0.02\\
 \hline
 PAN $(256\times256)$ \cite{C3} &  16.06 & 0.48 & 0.06 & 19.23 & 0.54 & 0.11 & 28.32 & 0.75 & 0.16\\
 \hline
 Pix2Pix-HD $(512 \times 512)$ \cite{C1} & - & - & - & 25.01 & \textbf{0.79} & \textbf{0.28} & \textbf{32.32} & \textbf{0.87} & \textbf{0.24}\\ 
 \hline
  Pix2Pix-HD $(2048 \times 1024)$ \cite{C1} & 22.18 & 0.66 & 0.14 & - & - & - & - & - & -\\ 
 \hline
 \textbf{MSG U-Net} $(512 \times 512)$ & - & - & - & \textbf{25.17} & 0.77 & 0.27 & 31.04 & 0.84 & 0.23\\ 
 \hline\textbf{MSG U-Net} $(2048 \times 1024)$ & \textbf{23.99} & \textbf{ 0.69} & \textbf{0.15} & - & - & - & - & - & -\\ 
 \hline
\end{tabular}
\label{tab:tab1}
\end{table*}

\section{Experiments and Results}
\label{sec:experimental_results}
We provide a comprehensive quantitative comparison against state-of-the-art methods over three datasets. 
% in Sec . \ref{subsec: quantitative_results}. 
% We then show results of our ablation studies on the MSG U-Net in Sec. \ref{subsec: ablations_studies}.
% \textbf{Implementation Details}: 
In all the experiments, we set the hyper-parameters $\alpha$ and $\beta$ described in equation [\ref{eq:lossfunc}] to $10$ and $0.25$, respectively, based on the extensive hyper-parameter tuning.
% \textbf{Evaluation Metrics}: 
We use Peak Signal-to-Noise Ratio (PSNR), Structural Similarity Index (SSIM) \cite{1284395}, and Visual Information Fidelity (VIF) \cite{1576816} as the evaluation metrics.
% \textbf{Compared Models}: 
We compare our method with the state-of-the-art image-to-image translation GANs, including Pix2Pix \cite{C2}, Perceptual Adversarial networks (PAN) \cite{C3}, and Pix2Pix-HD \cite{C1}.

% \textbf{Datasets Used}: 
We use three benchmark datasets, namely Cityscapes \cite{Cordts2016Cityscapes}, NYU Indoor RGBD \cite{Silberman:ECCV12}, and Aerial Images to Maps \cite{C2}.
Cityscapes dataset consists of segmented RGB image of city streets in the source domain and corresponding photo-realistic image in the target domain at resolution $2048\times1024$. 
% We set segmented image as source domain and photo-realistic image as target domain.
NYU Indoor RGBD dataset consists of segmented indoor images in the source domain and corresponding real images in the target domain at resolution $512\times512$. 
% We set segmented indoor images as source domain and real-images as target domain. 
Aerial Images to Maps dataset consists of aerial photographs in the source domain and corresponding map of the image in the target domain at resolution $512\times512$. 
% We set aerial photograph as source domain and map of photograph as target domain.

\begin{table}[!t]
\small
\caption{Ablation study on Cityscapes dataset \cite{Cordts2016Cityscapes}.  }
\centering
% \resizebox{\columnwidth}{!}{
\begin{tabular}{|c |c c c c c |} 
\hline
 Input & \multicolumn{5}{|c|}{Output Resolution}\\
 \hline
 Resolution &  128$\times$64  & 256$\times$128 & 512$\times$256 & 1024$\times$512 & 2048$\times$1024\\
 \hline
 128$\times$64 & 0.42  & 0.43 & 0.44  & 0.49 & 0.49 \\
 \hline
 256$\times$128  &  0.56 & 0.58 & 0.58 & 0.59 & 0.59\\
 \hline
 512$\times$256  & 0.59 & 0.63 & 0.63 & 0.63 & 0.63 \\ 
 \hline
 1024$\times$512 & 0.63 & 0.65 & 0.67 & 0.67 & 0.68 \\ 
 \hline
 2048$\times$1024 & 0.65 & 0.67 & 0.70 & 0.69 & 0.69\\
 \hline
\end{tabular}
% }
\label{tab:tab2}
\end{table}

\begin{table}[!t]
\centering
\caption{Ablation studies on  NYU Indoor RGBD \cite{Silberman:ECCV12} and Aerial Images to Maps \cite{C2} datasets. Dimension $S \times S$ is represented by $S$.}
% \resizebox{\columnwidth}{!}{
\begin{tabular}{|c |c c c c | c c c c|} 
\hline
 Input  & \multicolumn{4}{|c|}{Output over NYU} & \multicolumn{4}{|c|}{Output over Aerial} \\
  & \multicolumn{4}{|c|}{Indoor RGBD dataset} & \multicolumn{4}{|c|}{Images to Maps dataset} \\
 \hline
 Size &  64  & 128 & 256 & 512 & 64 & 128 & 256 & 512  \\
 \hline
 64 & 0.51  & 0.57 & 0.58  & 0.57  & 0.51  & 0.57 & 0.61  & 0.61\\
 \hline
 128 & 0.69 & 0.75 & 0.76 & 0.72 & 0.64 & 0.78 & 0.79 & 0.78\\
 \hline
 256  & 0.71 & 0.78 & 0.80 & 0.76 & 0.77 & 0.82 & 0.83 & 0.83\\ 
 \hline
 512 & 0.71 & 0.79 & 0.80 & 0.77 & 0.79 & 0.84 & 0.86 & 0.84\\ 
 \hline
\end{tabular}
% }
\label{tab:tab3}
\end{table}

\subsection{Quantitative Results}
\label{subsec: quantitative_results}
The experimental results in terms of the PSNR, SSIM and VIF are reported in Table \ref{tab:tab1} over Cityscapes, NYU Indoor RGBD, and Aerial Images to Maps datasets. Note that we remove the topmost input layer and the output layer of dimension $2048 \times 1024$ and resize the $1024 \times 512$ input and output resolution to $512 \times 512$ to facilitate the experiments on NYU Indoor RGBD and Aerial Images to Maps datasets.
It is noticed from the results that the proposed MSG U-Net outperforms the Pix2Pix, PAN, and Pix2Pix-HD models over Cityscapes dataset. However, the results of the MSG U-Net is very close to Pix2Pix-HD over NYU Indoor RGBD and Aerial Images to Maps datasets.
Note that Pix2Pix-HD is $2.5$ times computationally more expensive than our MSG U-Net.
Thus, the proposed model is very efficient as compared to the Pix2Pix-HD and generates the output images with similar quality.

\subsection{Ablation Study}
\label{subsec: ablations_studies}
The proposed MSG U-Net takes an input from the source domain and down-sample it to various resolutions.
In this experiment, we take the down-sampled input image such as $128 \times 64$ and again sample it to all input scales and feed it to generator.
% to observe the change in the quality of the generated image.
% generated by network when we use low resolution input. 
The purpose of this ablation study is to check the capability of MSG U-Net to perform both image-translation and super-resolution, simultaneously. We report SSIM under different input and output resolution settings in Table \ref{tab:tab2} over Cityscapes and Table \ref{tab:tab3} over NYU Indoor RGBD and Aerial Images to Maps datasets. 
It is observed that even if the resolution of input image is reduced roughly by half, the results are not much degraded. In fact, the results are mostly comparable for different input resolutions. This can be quiet useful during the inference if high-resolution input images are not available, we can still generate high-resolution target images.

\subsection{Qualitative Results}
\label{subsec:qualitative}
\begin{figure*}
     \centering
     \begin{subfigure}[b]{0.48\textwidth}
         \centering
         \includegraphics[width=\textwidth]{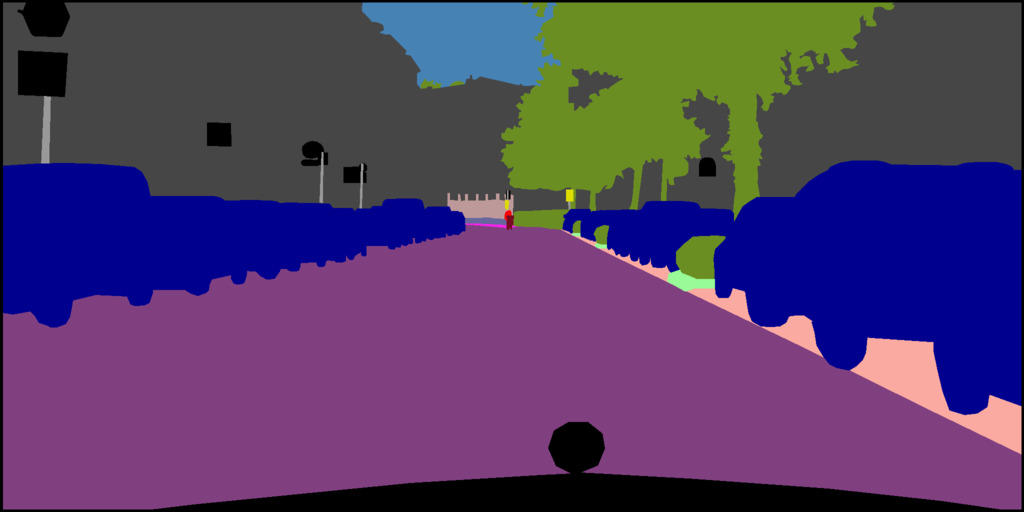}
          \caption{Source Image}
    \end{subfigure}
    \begin{subfigure}[b]{0.48\textwidth}
        \centering
        \includegraphics[width=\textwidth]{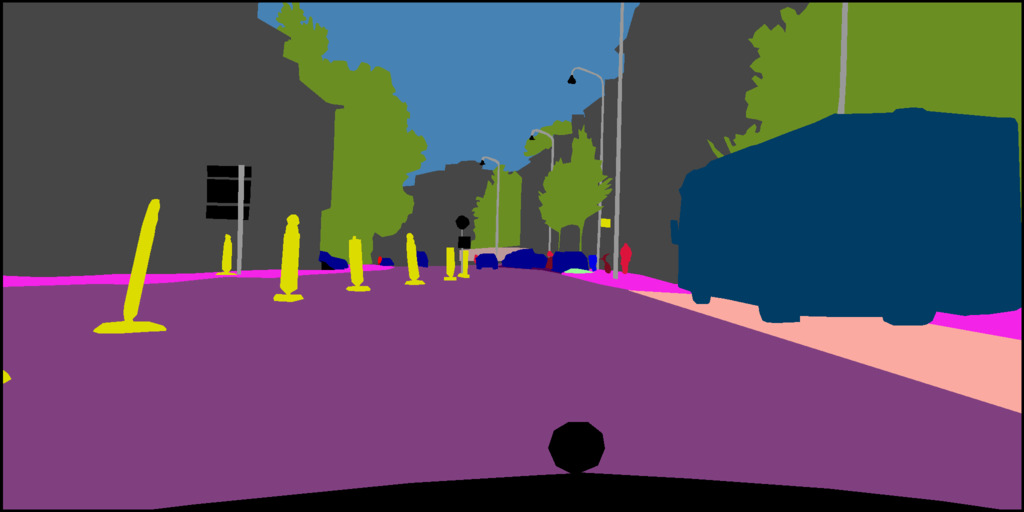}
        \caption{Source Image}
     \end{subfigure}\hfill
     \begin{subfigure}[b]{0.48\textwidth}
         \centering
         \includegraphics[width=\textwidth]{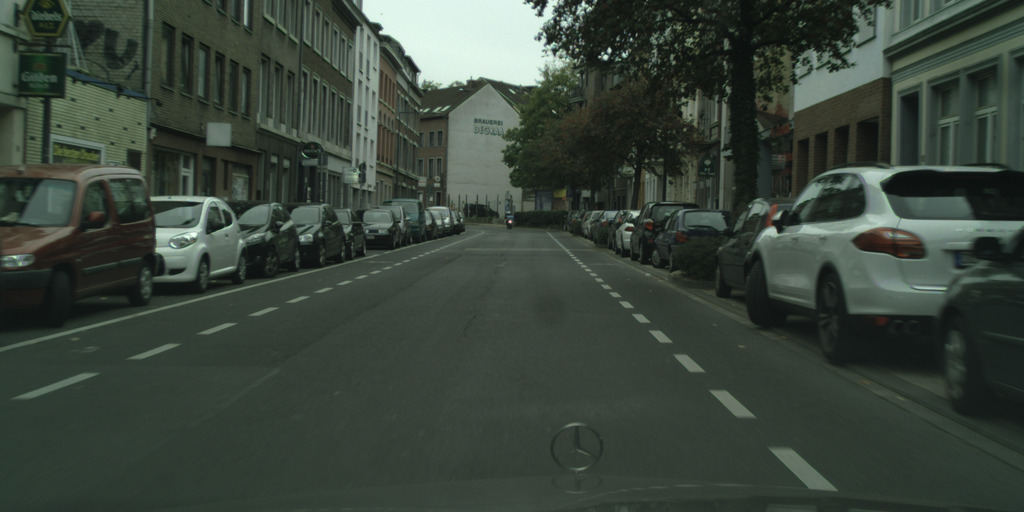}
          \caption{Target Image}
     \end{subfigure}
     \begin{subfigure}[b]{0.48\textwidth}
         \centering
         \includegraphics[width=\textwidth]{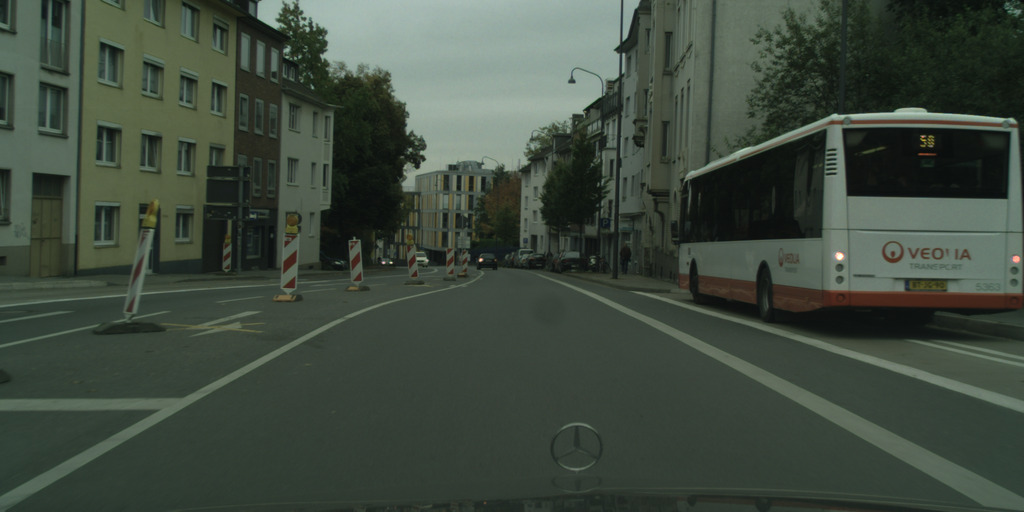}
          \caption{Target Image}
     \end{subfigure}
     \hfill
     \begin{subfigure}[b]{0.48\textwidth}
         \centering
         \includegraphics[width=\textwidth]{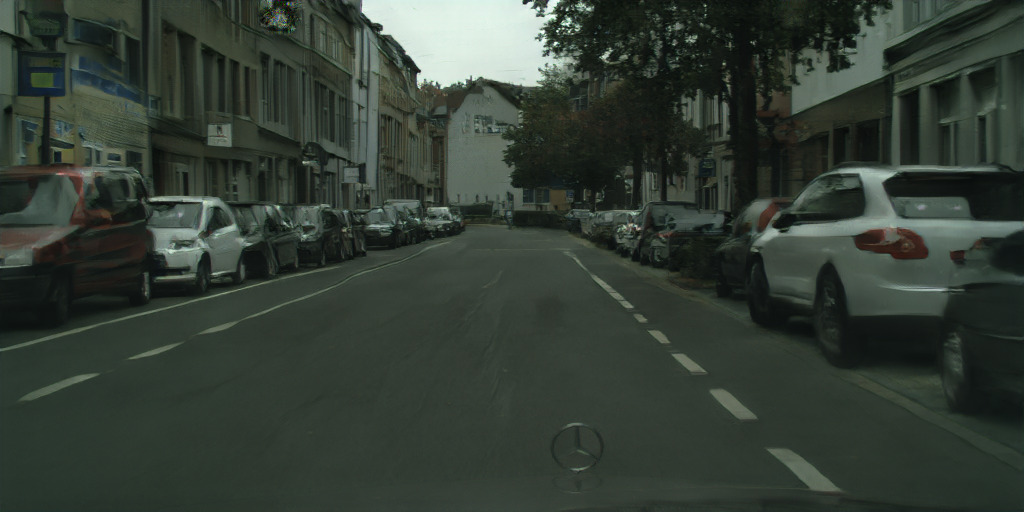}
         \caption{Generated Output}
     \end{subfigure}
     \begin{subfigure}[b]{0.48\textwidth}
         \centering
         \includegraphics[width=\textwidth]{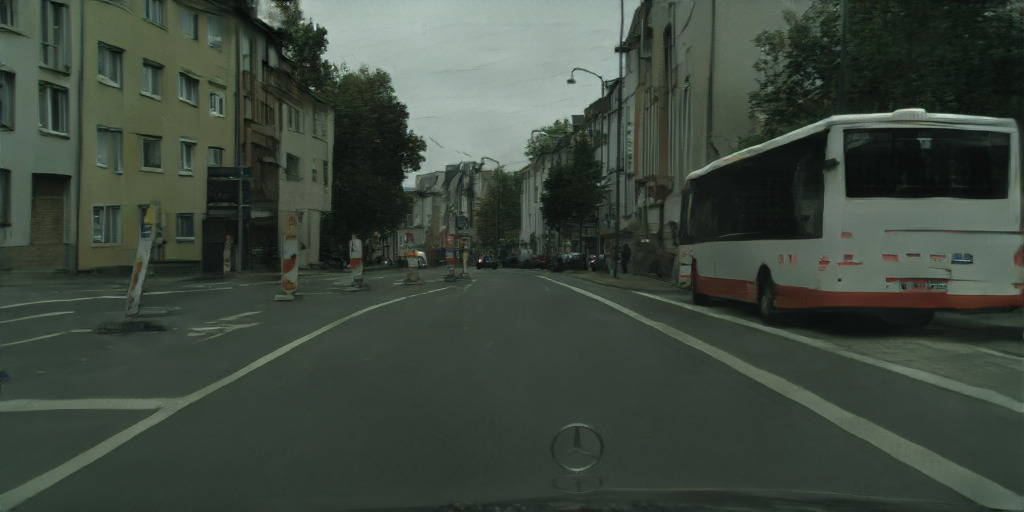}
         \caption{Generated Output}
     \end{subfigure}
     \caption{Example results on Cityscapes dataset. Here, we convert segmented-RGB image to photo realistic image.}
    \label{fig:citydata}
\end{figure*}

\begin{figure*}
    \centering
     \begin{subfigure}[b]{0.3\textwidth}
         \centering
         \includegraphics[width=\textwidth  ]{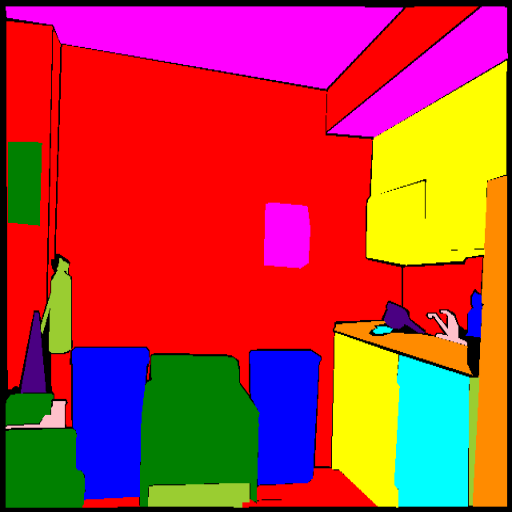}
         \caption{Input Image}
         
     \end{subfigure}
     \hfill
     \begin{subfigure}[b]{0.3\textwidth}
         \centering
         \includegraphics[width=\textwidth]{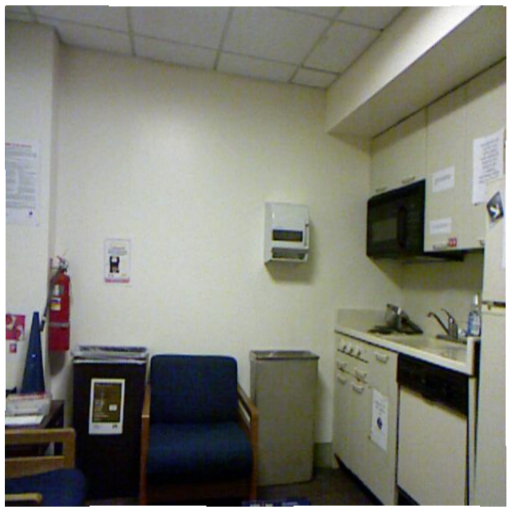}
         \caption{Target Image}
         
     \end{subfigure}
     \hfill
     \begin{subfigure}[b]{0.3\textwidth}
         \centering
         \includegraphics[width=\textwidth]{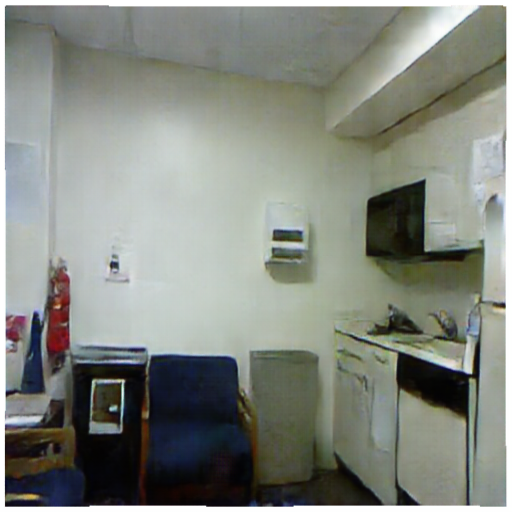}
         \caption{Generated Output}
     \end{subfigure}
     
      \begin{subfigure}[b]{0.3\textwidth}
         \centering
         \includegraphics[width=\textwidth]{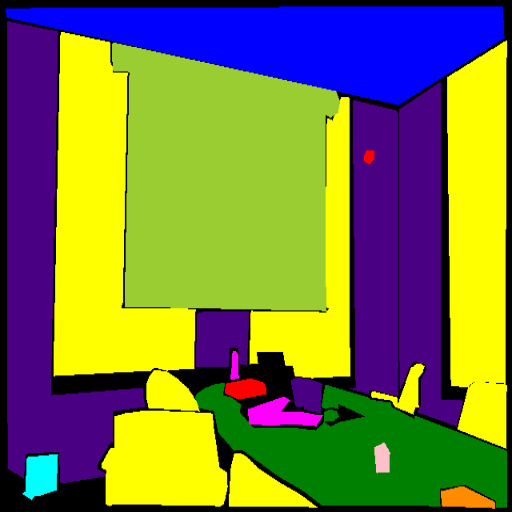}
         \caption{Input Image}
         
     \end{subfigure}
     \hfill
     \begin{subfigure}[b]{0.3\textwidth}
         \centering
         \includegraphics[width=\textwidth]{images/27_target.png}
         \caption{Target Image}
         
     \end{subfigure}
     \hfill
     \begin{subfigure}[b]{0.3\textwidth}
         \centering
         \includegraphics[width=\textwidth]{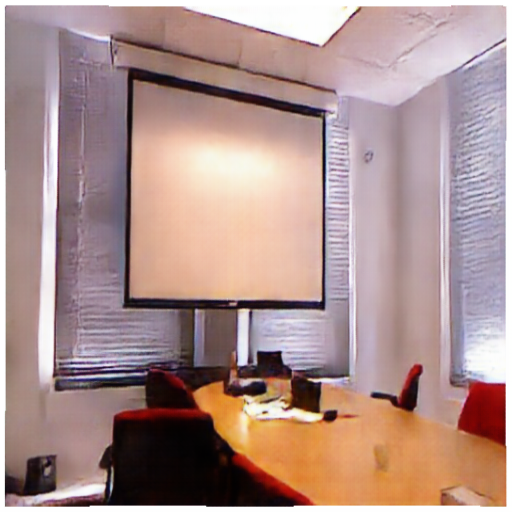}
         \caption{Generated Output}
     \end{subfigure}
     
           \begin{subfigure}[b]{0.3\textwidth}
         \centering
         \includegraphics[width=\textwidth]{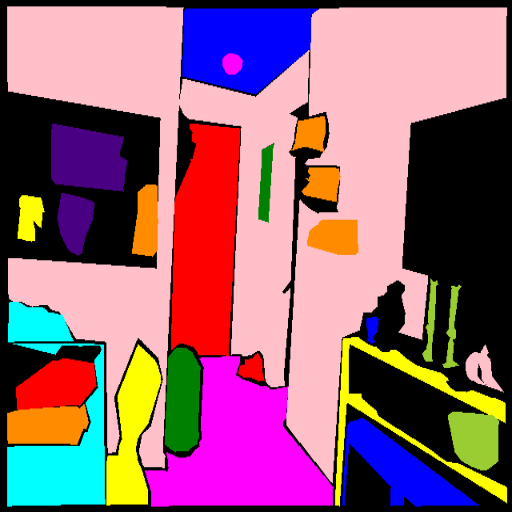}
         \caption{Input Image}
         
     \end{subfigure}
     \hfill
     \begin{subfigure}[b]{0.3\textwidth}
         \centering
         \includegraphics[width=\textwidth]{images/53_target.png}
         \caption{Target Image}
         
     \end{subfigure}
     \hfill
     \begin{subfigure}[b]{0.3\textwidth}
         \centering
         \includegraphics[width=\textwidth]{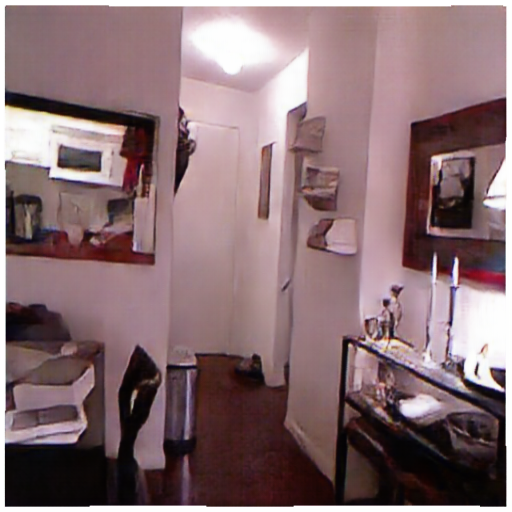}
         \caption{Generated Output}
     \end{subfigure}
    \caption{Example results on NYU Indoor RGBD dataset images. Here, we convert segmented indoor images to photo realistic images.}
    \label{fig:nyu_data}
\end{figure*}

\begin{figure*}
    \centering
     \begin{subfigure}[b]{0.3\textwidth}
         \centering
         \includegraphics[width=\textwidth  ]{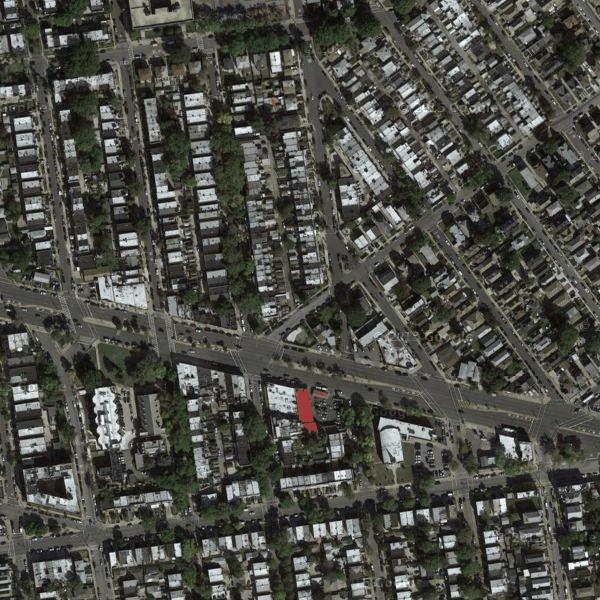}
         \caption{Input Image}
         
     \end{subfigure}
     \hfill
     \begin{subfigure}[b]{0.3\textwidth}
         \centering
         \includegraphics[width=\textwidth]{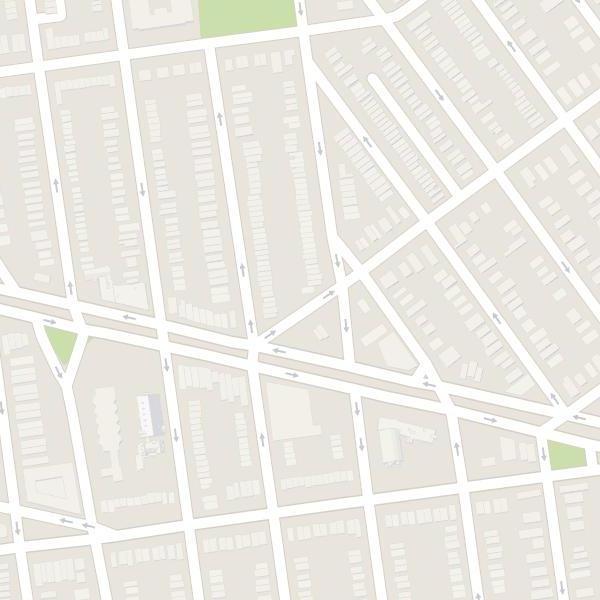}
         \caption{Target Image}
         
     \end{subfigure}
     \hfill
     \begin{subfigure}[b]{0.3\textwidth}
         \centering
         \includegraphics[width=\textwidth]{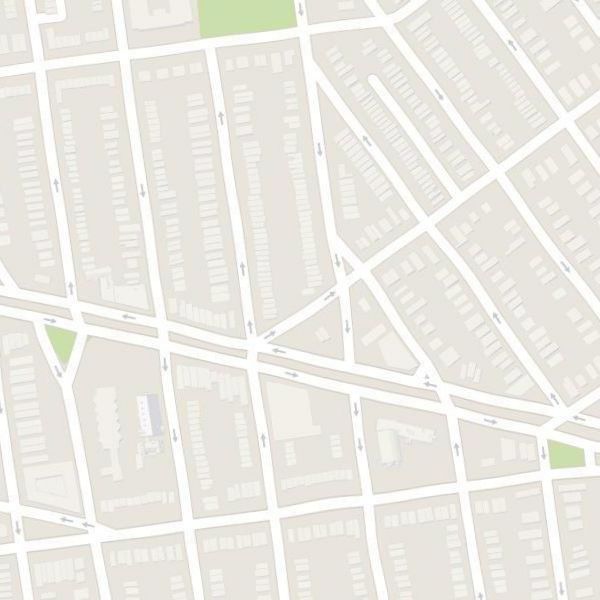}
         \caption{Generated Output}
     \end{subfigure}
     
      \begin{subfigure}[b]{0.3\textwidth}
         \centering
         \includegraphics[width=\textwidth]{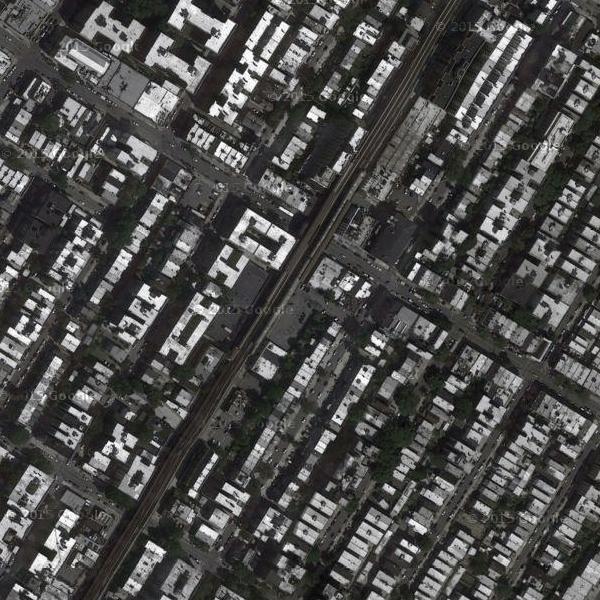}
         \caption{Input Image}
         
     \end{subfigure}
     \hfill
     \begin{subfigure}[b]{0.3\textwidth}
         \centering
         \includegraphics[width=\textwidth]{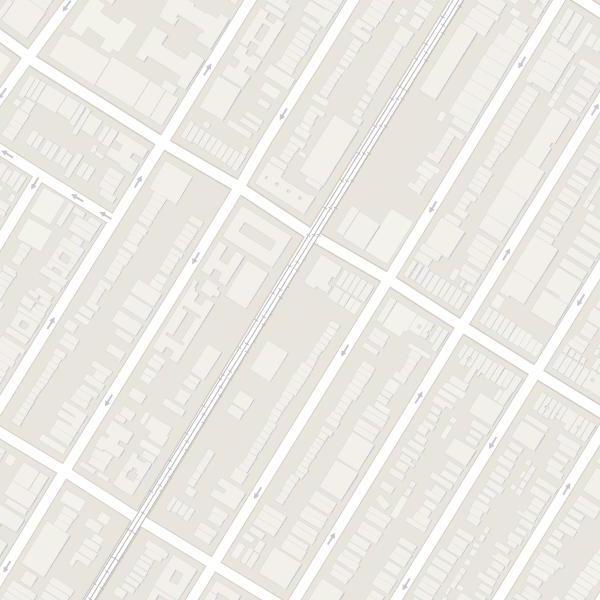}
         \caption{Target Image}
         
     \end{subfigure}
     \hfill
     \begin{subfigure}[b]{0.3\textwidth}
         \centering
         \includegraphics[width=\textwidth]{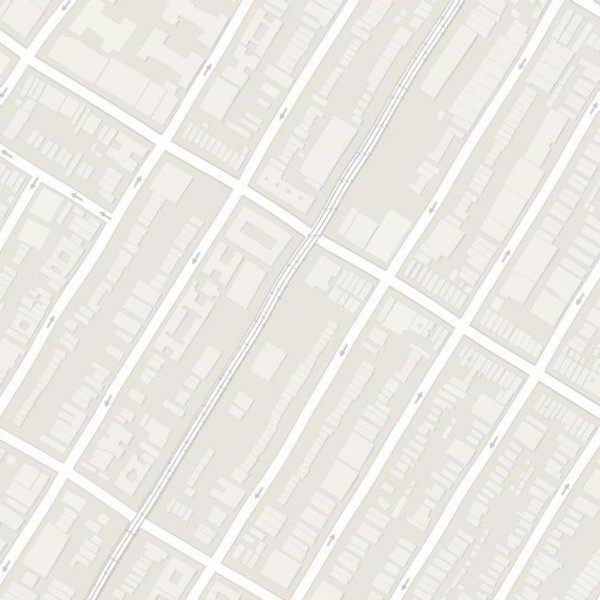}
         \caption{Generated Output}
     \end{subfigure}
     
           \begin{subfigure}[b]{0.3\textwidth}
         \centering
         \includegraphics[width=\textwidth]{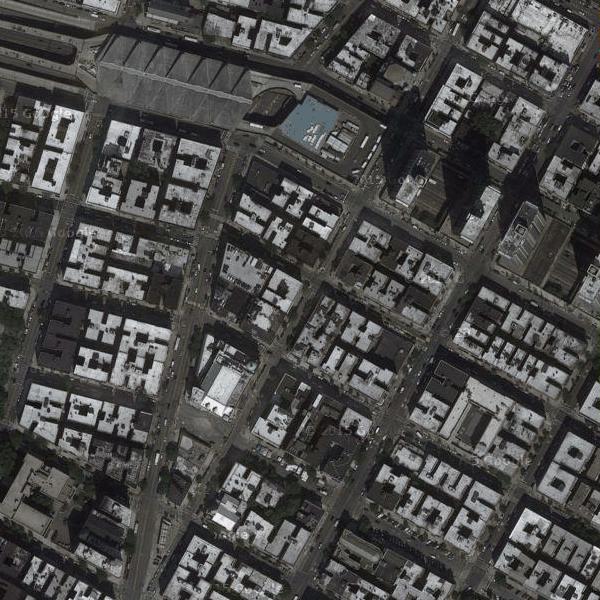}
         \caption{Input Image}
         
     \end{subfigure}
     \hfill
     \begin{subfigure}[b]{0.3\textwidth}
         \centering
         \includegraphics[width=\textwidth]{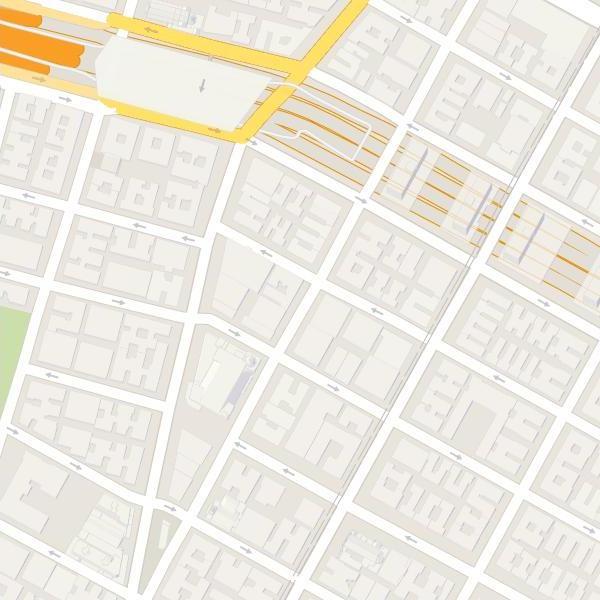}
         \caption{Target Image}
         
     \end{subfigure}
     \hfill
     \begin{subfigure}[b]{0.3\textwidth}
         \centering
         \includegraphics[width=\textwidth]{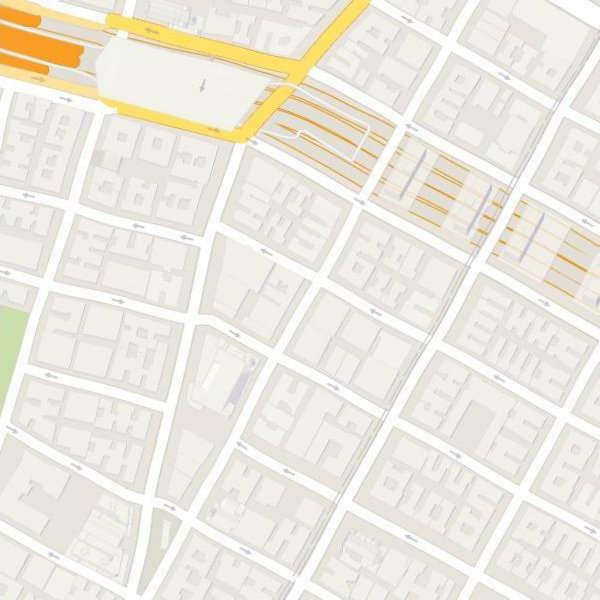}
         \caption{Generated Output}
     \end{subfigure}
    \caption{Example results on Aerial Images to Maps dataset images. Here, we convert aerial photographs to maps.}
    \label{fig:map_data}  
\end{figure*}

We present the images generated by the proposed MSG U-Net model in this section. To be precise Fig. \ref{fig:citydata} shows the source images, target images and generated images from Cityscapes dataset \cite{Cordts2016Cityscapes}. Fig. \ref{fig:nyu_data} shows results on NYU Indoor RGB-D dataset \cite{Silberman:ECCV12}. Fig. \ref{fig:map_data} shows results on Aerial Images to Maps dataset \cite{C2}. The quality of the generated images can be easily observed for the high-resolution synthesis in these results.

% \begin{figure*}
%      \centering
%      \begin{subfigure}[b]{\textwidth}
%          \centering
%          \includegraphics[width=\textwidth]{images/aachen_000062_000019_input.jpg}
%           \caption{Source Image}
%      \end{subfigure}
%      \hfill
%      \begin{subfigure}[b]{\textwidth}
%          \centering
%          \includegraphics[width=\textwidth]{images/aachen_000062_000019_target.jpg}
%           \caption{Target Image}
%      \end{subfigure}
%      \hfill
%      \begin{subfigure}[b]{\textwidth}
%          \centering
%          \includegraphics[width=\textwidth]{images/aachen_000062_000019.jpg}
%          \caption{Generated Output}
%      \end{subfigure}
%      \caption{Example results on Cityscapes dataset. Here, we convert segmented-RGB image to photo realistic image.}
%      \label{fig:C_fig2}
% \end{figure*}

\section{Conclusion}
\label{sec:conclusion}
In this paper, we proposed a Multi-scale gradient based U-Net (MSG U-Net) for high-resolution image-to-image translation. The proposed model shows better training due to the utilization of the gradients at different scales. 
% The results are also computed and compared with state-of-the-art models over three benchmark datasets.
The proposed MSG U-Net (1.28 Tera-Flops) is computationally efficient as compared to Pix2Pix-HD (3.32 Tera-Flops). In spite of being an efficient model, the MSG U-Net has shown either better or comparable performance over three benchmark datasets. It is also observed that the MSG U-Net can produce high-quality output images even from the low-resolution input images. 
% The ablation studies shows that even if the resolution of input image is low, the quality of the images generated by the model remains similar which shows that MSG U-Nets can be robust and effective, even if the resolution of input images are lowered.

\bibliographystyle{splncs04}
\bibliography{References}




% LNCS Homepage, \url{http://www.springer.com/lncs}. Last accessed 4
% \end{thebibliography}
\end{document}